\documentclass[aps,numerical,graphicx,showpacs,prl]{revtex4-1}
\usepackage{amsmath}
\usepackage{graphicx}
\draft 

\begin{document}
%
%
\def\Journal#1#2#3#4{{#1 }{\bf #2, }{ #3 }{ (#4)}}

\title{
Breaking news on last achievements  on the definition of the black-body total internal energy} 
\author{Lino Reggiani}
\email{lino.reggiani@unisalento.it}
\affiliation{Dipartimento di Matematica e Fisica, ``Ennio de Giorgi'',
Universit\`a del Salento, via Monteroni, I-73100 Lecce, Italy}

\author{Eleonora Alfinito}
\affiliation{Dipartimento di Matematica e Fisica, ``Ennio de Giorgi'',
Universit\`a del Salento, via Monteroni, I-73100 Lecce, Italy}

\date{\today}

\begin{abstract}
The internal total-energy of the black-body is a physical quantity of paramount importance in the development of modern physics. 
Accordingly, together with a brief historical development, we report and comment last breaking news (2018-2024)  concerning the definition and properties of this quantity.
The first comment concerns with the inclusion of the Casimir energy that avoids the vacuum catastrophe implied by he presence of zero-point energy, thus leading to further quantum contributions associated with boundary effects.
The second comment concerns with  a semi-classical simulation of a one dimensional black-body whose results suggest a possible reconsideration on the role of classical physics on the quantum black-body.

\end{abstract}
%


\maketitle 

%
\section{Introduction} 
The internal energy of a thermodynamic system is a fundamental quantity to describe its macroscopic properties on the basis of the law of energy conservation  and thermodynamic transformations. 
The first system investigated  thermodynamically was the ideal massive gas.
Then, the development of statistical physics provided the scientific method to derive the proper microscopic interpretation of others  macroscopic physical systems.
In particular, the second system investigated  was the ideal electromagnetic (EM) gas that was used to model the interaction between matter and light, specifically the black-body system.
The black-body originates from the Stefan law suggested in 1874 \cite{stefan,crepeau09} in a first attempt to fit existing experiments concerning the radiation emitted by a body in thermal equilibrium conditions at a given temperature $T$. 
 In its simplest form, at thermal equilibrium  with its environment  the black-body internal energy, $U$,  is instantaneously  equal to the radiated-energy and given by:
\begin{equation}
 U = cons \times T^4
\label{e1a}
\end{equation}
where $cons$ is a constant keeping the dimensionality of Eq. (\ref{e1a}) to be fitted by experiments (as done empirically by Stefan in 1874) or by an appropriate microscopic theory (as started  by Bolzmann in 1884).  
\par
The formulation of a microscopic theory by Boltzmann  is still a source of debate, as summarized in Table 1, and briefly discussed in the following sections. 
\par
\begin{table}[pt]
\caption{Brief historical overview of the internal energy $U$ of a black-body with volume $V$.}
{\begin{tabular}{@{}ccccc@{}} \toprule \\
		Year & Author & Contribution & Physical approach & Consequences \\
		\colrule
		1884 & Boltzmann  & $U^B= CVT^4$& Thermodynamics & $C$ from experiments \\
		1900 & Rayleigh-Jeans  & $U^{RJ} = lim_{f \rightarrow \infty}
		\frac{8 \pi f^2}{c} K_BT = \infty$ & Classical statistics & Ultraviolet catastrophe\\
		1901 & Planck & $ U^P= \frac {2 \pi^5} {15c^3 h^3}  V (K_BT)^4 $ & Quantum statistics & $ C= \frac{2 \pi^5 K_B^4}{15 c^3 h^3}$  \\
		1912 & Planck & $U^{P1}=U^P+ 1/2\sum hf$ & Quantum statistics + zero-point energy & Vacuum catastrophe\\
		1951 & Callen-Welton & $U^{CW}=U^P + 1/2\sum hf$ & Quantum perturbation theory & Vacuum catastrophe \\
		1957 & Kubo & $U^K=U^P+ 1/2\sum hf$ & Quantum correlation-response formalism & Vacuum catastrophe \\
		2018 & Reggiani Alfinito & $U^{RA}=U^P + U^{ZP, Casimir}$ & Quantum statistics + Casimir energy & Radiation corrections\\
		2022 & Wang Casati Benenti & $U^{WCB} = C'T^2$ & 1D classical Hamiltonian simulation & 1D classical S-B law \\
	\botrule	
		\end{tabular}}
\label{tab:BriefHistoricalOverview}
\end{table}
\section{Boltzmann thermodynamic model} 
Boltzmann in (1884)
justified the derivation of the empirical Stefan law  by using thermodynamics applied to a not well defined electromagnetic (EM) gas described by Maxwell field equations where: (i) the internal energy $U$was postulated  to be function of volume $V$ and temperature $T$ as:   
\begin{equation}
U=U(V,T)
\label{e1}
\end{equation}
and, (ii) following Maxwell equations, the state equation of the EM gas was given by
\begin{equation}
U=3 PV
\label{e2}
\end{equation}
with $P$ the pressure exerted by the EM fluid on the walls of a thermodynamic container of volume $V$ at a given  temperature $T$.
\par
A further  implicit assumption  was that, for finite values of  $V$ and $T$, also $U$ takes finite values.
Then,  by using first and second thermodynamics principles adequate to a canonical ensemble, Boltzmann obtained the empirical Stefan law in he form
\begin{equation}
U^{SB}
=CVT^4
\end{equation}
with the radiation density constant  $C=cns V£$ an integration constant which value could be determined by fitting with experiments as:
\begin{equation}
C= 7.57 \times 10^{-16}   \ \  [J m^{-3}K^{-4}]
\end{equation}
\section{Rayleigh-Jeans classical Maxwell model} 
Rayleigh-Jeans \cite{r00,j00}, by adopting a classical statistics approach, considered an ideal black-body with metallic walls and filled by an EM gas consisting of an ensemble of normal modes satisfying Maxwell equations in vacuum, that is far from the EM sources \cite{leff02}.
Then, the energy dispersion of each normal-mode, 
$\epsilon_{max}$, is given by
\begin{equation}
\epsilon_{max}=pc
\label{e3}
\end{equation}
where the $max$ subscript stands for Maxwell-modes,  being $p$ the modulus of the EM momentum, as given by the Poynting vector \cite{jackson62}  of each single mode.
Notice, that even if the Maxwell mode is an harmonic function of space and time (a typical plane wave) its energy is a continuous function of the EM momentum amplitude only and, being  independent of the EM frequency and wave-vector, according to the Rayleigh-Jeans model \cite{r00,j00},  becomes responsible for a divergence of the  total-energy $U$ inside the cavity in the limits of an infinite number of Maxwell modes at increasing frequencies. This is a condition compatible with Maxwell equations,  also  known as ultraviolet catastrophe
\cite{kafri19}, in complete disagreement with experiments and SB law.
\par
We stress that, contrary to what is usually reported in the literature, here we consider Maxwell equations already pertaining to modern physics (i.e. relativity) and not to classical physics strictly pertaining to Newton and the rational mechanics mostly developed in the period between Newton and Maxwell discoveries.
\par
The space-time average energy per mode, $<\epsilon_{cm}>$ in the case of classical statistics is \cite{webpg1}:
%
\begin{equation}
<\epsilon_{cm}> = \frac{V (E^2 + B^2)}{8 \pi} = K_BT
\label{e2a}
\end{equation}
with $E$ and $B$ the maximum amplitude of the electric and magnetic fields, respectively, and where the second equality represents the classical energy-equipartition law \cite{kafri19}.
\par
The number of classical modes per unit frequency and unit 
volume $N_c$, is given by 
\begin{equation}
N_c(f) = \frac{8 \pi f^2}{c^3}
\label{e3a}
\end{equation}
with $f$ the mode frequency, as amply documented in the open literature \cite{webpg1} 
The product of the above two quantities, $N_c(f)$ and $K_BT$, represents the Rayleigh-Jeans \cite{r00,j00} approximation formula for the internal energy of the black-body at frequency $f$
\begin{equation}
U^{RJ}(f) = N_c(f) K_BT
\label{e4a}
\end{equation}
that, by construction,  is consistent with Maxwell equations and classical statistics and, in the limit $f \rightarrow \infty$, predicts the ultraviolet catastrophe.
(In the recent literature \cite{wang22} it is mentioned that the ultraviolet catastrophe of the Rayleigh-Jeans model can be avoided since equipartition is not strictly valid because ergodic conditions are not satisfied for an infinite number of modes.  Since here we are interested in the general increasing trend of $N(f)$ this ergodic problematic does not modify the main conclusions of the  Raleigh-Jeans model.) 
\section{First Planck model}
With the birth of quantum mechanics, announced by Planck in 1901 \cite{planck01},
the classical normal-modes were replaced by the quantum normal-modes, known as  photons, as coined by Lewis in 1926 \cite{lewis26}, with energy  depending linearly on frequency and quantized by Planck as:
\begin{equation}
\epsilon_p=n hf 
\label{planck1}
\end{equation}
where the $p$ subscript stands for photons being $n$ an integer number.
Accordingly,  the average value of the internal-energy is found to  agree with the implicit form of Eq. (\ref{e1}) and explicitly given by \cite{reggiani22}:
\begin{eqnarray}
\overline U^{P1}(V,T) &=&  \frac {8 \pi \Gamma(4) \zeta(4)} {c^3 h^3}  V (K_BT)^4 \nonumber \\ 
&=&  C V T^4 
\label{e9}
\end{eqnarray}
with $\Gamma(4) \zeta(4) = \pi^4  /  15
=6.49$.
\par 
Further property of the photon gas inside the black-body was further investigated on the basis of Bose-Einstein quantum statistics and found that photons  instantaneous number is not defined, but its average number is finite and depends on temperature and volume as given by \cite{reggiani22,leff02}: 
\begin{eqnarray}
\overline N^P(V,T)  &=&  \frac{8 \pi \Gamma(3) \zeta(3) }{c^3 h^3}\ V (K_BT)^3 \nonumber \\ 
 &=&  (2.02 \ 10^7)  \ V T^3    
\label{e10}
\end{eqnarray}
with  $\Gamma(3) \zeta(3) =\int_0^{\infty} x^2/(e^x-1) dx = 2.404$, being 
$\Gamma$ and $\zeta$ respectively the Gamma and the Riemann functions.
\par
Accordingly,  one can write:
\begin{equation}
\overline \epsilon_{p} = \frac{\overline U^P} {\overline N^P} 
= \ 2.7 \  K_BT
\label{e11}
\end{equation}
with $\overline{\epsilon_{p}}$ being the average energy per photon mode.
\par
We notice that, because of quantum statistics, the numerical value of $2.7$ is slightly less than the value of $3$ pertaining to the classical case  per full relativistic massive particles. 
Among the predictions of the quantum statistics there were the microscopic derivation of the SB law and the determination of the radiation constant given in terms of universal constants in perfect agreement with Eq. (\ref{e9}).
\par
We want to stress,  that
the explicit theoretical value of the radiation constant comes out to agree quite well with the experimental value, without  forcing the fit with experiments. In this way, also the ultraviolet catastrophe was avoided, even if in the classical limit $h \rightarrow 0$ the radiation constant of the quantum  SB law tends to infinite  thus recovering the ultraviolet catastrophe predicted by classical statistics.
\section{Second Planck model}
By ignoring the symmetry between emission and absorption, Planck maintained that the absorption of radiation energy is continuous.
Under these assumptions, Planck derived in 1912 \cite{planck12} a second
radiation law by including a zero-point energy contribution \cite{dannon05}.
Accordingly, $n$  was  replaced $n + 1/2$ in Eq. (\ref {planck1}) thus leading to the possibility of  a vacuum catastrophe \cite{adler95}. 
Indeed, the zero-point contribution when calculated over the total number of photons in vacuum leads to:
\begin{equation}
U^{P2} =  U^{P1} +  U^{ZP}
\end{equation}
with
\begin{equation}
U^{ZP} =  \sum \frac{hf}{2} = \infty
\end{equation}
We remark that Planck did not commented on this catastrophe possibility.
We will come back to this point in the next sections. 
\section{Callen-Welton and Kubo  quantum model with zero point}
Fundamental advances based on the advance of quantum mechanics  started from Callen and Welton 
(1951) \cite{callen51}, where a first order perturbation theory was used, and Kubo operatorial formalism of 1957 \cite{kubo66}.
We remark that both the approaches confirmed the emergence of the zero-point energy contribution  that
originated a debate by the scientific community about the correctness or less of the derivation and the experimental evidence of the zero-point energy. 
For more details on the subject the reader is sent to the next section of the paper. 
\section{Reggiani-Alfinito full quantum model}
Reggiani-Alfinito \cite{reggiani18} by considering Casimir effect \cite{casimir48}, associated with the presence of ideal opposite  metallic walls in the black-body  responsible to avoid vacuum catastrophe due to zero point contribution,y modified the expression of the total internal energy now expressed as:
\begin{equation}
U^{RA}=U^{P1} + U^{ZP,Casimir} 
\end{equation}
with the Casimir energy given by:
\begin{equation}
U^{ZP,Casimir} 
=- \frac{\pi h c A}{1440 L^3} 
\label{eq19} 
\end{equation}
Accordingly, the vacuum catastrophe is avoided by the Casimir effect which is responsible for an energy correction to the total-energy value in general negligible.   
\par
We can go further in revisiting the SB law by noticing the presence of quantum effects in terms of the existence of 
a radiation pressure, $p_{R}$, exerted by the photon gas inside the black-body cavity on the wall of the cavity, and of a
a Casimir pressure, $p_{C}$, associated with the presence of zero-point energy contribution, and whose definitions and physical meaning are briefly explained below.
\par
The radiation pressure is exerted by the photon gas on a given internal wall of the cavity, and is a function of absolute temperature only and according to Planck radiation law is given in Pascal $[Pa]$ by:
\begin{equation}
p_{R}= \frac{4L \sigma}{c}T^4
= \frac{8 \pi^5 K_B^4}{45 h^3 c^3} T^4
=49.8 \times 10^{-16} \ T^4  \ [Pa]
\end{equation}
The Casimir pressure is exerted by the parallel metallic plate distant a length  $L$ from each other,  it is a function of the distance $L$, and is given in Pascal $[Pa]$ by:
\begin{equation}
p_{C}= -\frac{\pi  h c}{480}L^{-4}
=-0.130 \times 10^{-26} \  L^{-4} \ [Pa]
\end{equation}
The minus sign indicates  that the Casimir pressure is exerted from the outside to the inside of the cavity. 
\par
We remark the need to introduce a reaction pressure $p_{rea}$,  that is necessary to keep the stability of the black-body cavity.
The reaction pressure necessary to balance the Casimir pressure to keep stability  conditions of the black-body cavity should be equal and opposite to the Casimir pressure and thus given by:
\begin{equation}
p_{rea}= -p_{C} 
\end{equation}
By considering that typical values of the pressure for a metallic structure to deviate from elastic conditions and/or of the temperature to avoid melting conditions of the structure are of the order of $10^{11} \ Pa$ and/or $10^3 \ K$, respectively  we conclude that instability conditions are expected for sizes below about nanometer length  or temperatures over about $10^3 \ K$. 
Under these conditions, the SB law no longer applies.
To this purpose, new experiments for the determination of the emission power under low temperatures and/or  small size black-body cavity should confirm the expectations given in the present paper.
\section{Wang Casati Benenti classical 1D  Hamiltonian model}
Wang et al \cite{wang22} by a  simulation technique  using a Newton-Maxwell coupling re-opened the interplay between classical/quantum interpretation of the SB law. 
Their Hamiltonian model, by coupling  classical Newton  mechanics with relativistic Maxwell equations, is not Lorenz invariant and therefore not physically justified. As a consequences, apart  the numerical simulations, no  reliable physical results can be extracted. 
We simply notice that they considered the Boltzmann derivation of the Stefan law  as obtained on a pure classical thermodynamics basis.  
As such, they erroneously concluded that Stefan-Boltzmann law, and more generally the spectral law of radiation emission, might be a consequence of the classical equations of motion.
On this subject it is worth mentioning a publication by Paul et al \cite{paul15} of 2015 with the provoking title:  
The Stefan–Boltzmann law: two classical laws (i.e. thermodynamics and electrodynamics) give a quantum one.
\section{Conclusions}
Looking back at the historic evolution of the black-body  total internal energy we list the following main conclusions.

1 - The main step was the Boltzmann formulation in 1884 of Stefan law which contained the fundamental constant of modern physics: $K_B$, $c$ and $h$. The magic intuition was the formulation of the form $U=U(V,T)$ later confirmed by Planck in 1901 with his EM energy quantization. On this ground, we believe that the usual claim in the literature that Boltzmann derivation follows from pure classical physics is not correct. Boltzmann used Maxwell equations to formulate the state equation of the EM gas, and Maxwell equations contains basically relativistic concepts,  not compatible with Newton classical mechanics.  Furthermore, the assumption of the form $U(V,T)$ is not compatible with any massive classical-gas.  Therefore, to our opinion a description of a black-body on the basis of Maxwell and Newton equations only is physically not-compatible. From our point of view classical physics starts with Newton and ends with the formulation of Maxwell electrodynamics equations, that already contain relativistic principles. Then with Maxwell, Planck, Einstein and advanced quantum mechanics we assist to the advent of Modern Physics.
  
2 - With Planck 1912, Callen-Welton 1951, kubo 1956 we assist to the development of the black-body description in terms of advanced quantum mechanics, including the problem of zero-point energy contribution. 

3 -  With Reggiani-Alfinito, 2018 the zero-point contribution was assimilated to the Casimir effect, introducing the so-called radiation correction associated with a finite-size effect due to the geometry and the material of the black-body cavity which is still open to further investigations.

4 - Classical problems related to ergodicity  and/or energy-equipartition relaxation Wang et al (2022) concerning the time evolution of simple Newton-Maxwell Hamiltonian systems attracted some attention without producing new or reliable physical results.

5 - The inclusion of Casimr effects is of fundamental importance since it avoids the vacuum catastrophe associated with zero-point energy contribution. However, at present calculations of the Casimir energy is limited to the case of an ideal metallic environment.  In general, calculations and experiments  concerning other materials and eventually other geometries for the black-body, and the associated Casimir energy needs further investigations  from both a theoretical and experimental side.

\end{document}